\documentclass{aa}

\usepackage{graphicx}
\usepackage{natbib}
\usepackage{pifont}

\usepackage[shortlabels]{enumitem}
\usepackage{amsmath}    
\usepackage{amssymb}    

\begin{document}

\title{Fluorine in the Solar Neighborhood: Chemical Evolution Models” }

\author {E. Spitoni \inst{1,2} \thanks {email to:
    spitoni@oats.inaf.it} \and F. Matteucci\inst{1,2,3} \and H. J\"onsson \inst{4} \and N. Ryde \inst{4} \and D. Romano \inst{5}} \institute{ Dipartimento di Fisica, Sezione di Astronomia,  Universit\`a di Trieste, via
  G.B. Tiepolo 11, I-34131, Trieste, Italy \and I.N.A.F. - Osservatorio
  Astronomico di Trieste, via G.B. Tiepolo 11, I-34131, Trieste,
  Italy \and I.N.F.N. -  Sezione di Trieste, Via Valerio 2, I-34100 Trieste, Italy \and Lund Observatory, Department of Astronomy and Theoretical Physics, Lund University, Box 43, SE-22100 Lund, Sweden \and I.N.A.F. - Osservatorio Astronomico di Bologna, Via Gobetti
93/3, I-40129 Bologna}

\date{Received xxxx / Accepted xxxx}

\abstract {In the light of the new observational data related to fluorine abundances in the solar neighborhood stars, we present here chemical evolution models testing different fluorine nucleosynthesis prescriptions with the aim to  best fit those new data.  }{We consider chemical evolution models in the solar neighborhood
testing different nucleosynthesis prescriptions for fluorine production  with the aim of reproducing the observed abundance  ratios [F/O] vs. [O/H] and [F/Fe] vs. [Fe/H]. We study in detail the effects of different stellar  yields on the fluorine production.}{The adopted chemical evolution models are: i) the classical ``two-infall'' model which follows the chemical evolution of halo-thick disk and thin disk phases, ii)  and the `` one-infall''  model designed only for  the thin disk evolution. We tested the effects on the predicted fluorine abundance ratios of different  nucleosynthesis yield sources:  AGB stars, Wolf-Rayet stars,  Type II and  Type Ia supernovae, and novae. } {The   fluorine production is dominated  by AGB stars  but the Wolf-Rayet stars are required to  reproduce the trend of  the observed data in the solar neighborhood by J\"onsson et al. (2017a) with our chemical evolution models. In particular, the best model both for the ``two-infall'' and ``one-infall'' cases requires an increase by a factor of two  of the Wolf-Rayet yields given by Meynet \& Arnould (2000).  We also show that the novae, even if their yields are still uncertain, could help to better reproduce the secondary behavior of F in the [F/O] vs. [O/H] relation. }{The inclusion of the fluorine production by Wolf-Rayet
stars seems to be essential to reproduce the  observed ratio  [F/O] vs [O/H] in the solar neighborhood
by  J{\"o}nsson et al. (2017a). Moreover, the inclusion of novae helps substantially to reproduce the observed fluorine secondary behavior. }


\titlerunning{Fluorine in the Solar neighborhood}
\authorrunning{Spitoni et al.}
\maketitle

\section{Introduction}

The aim of this paper is to study the evolution of fluorine ($^{19}$F)
abundance in the solar neighborhood, by means of a detailed chemical
evolution model.
The origin of fluorine is still uncertain and widely debated in
literature.  The production of the only stable isotope, $^{19}$F, is
strictly linked to the physical conditions in stars. In particular, we
can identify several stellar sites for fluorine production:
\begin{enumerate}[i)]
\item Asymptotic giant branch (AGB) stars. For solar metallicity the main
process of $^{19}$F production is related to nuclear reaction on
$^{14}$N including $\alpha$, neutron, and proton captures.  On the
other hand, at lower metallicity the $^{19}$F production depends on
$^{13}$C (Cristallo et al. 2014).
\item Wolf-Rayet (W-R) stars can be
important producers of fluorine which is injected into ISM by their
strong stellar winds. Also here, the chain of reactions leading to
$^{19}$F production starts from $^{14}$N, which is normally produced
during the CNO cycle as a secondary element. However, in massive stars
N can be produced as a primary element if they suffer strong rotation,
which produces effects similar to those to the dredge-up mechanism in
AGB stars. In this case, however, the primary N production is
limited only to very metal poor stars  (Z$<10^{-5}$) as shown by
Meynet \& Maeder (2002). The difference secondary/primary N is
important, since in the case of secondary N the F behavior would also
follow that of a secondary element, namely depending on the original
stellar metallicity.
\item Type II supernovae (SNe II). These SNe can
produce  $^{19}$F via the neutrino ($\nu$) process. In fact,
although  neutrinos are characterized by small cross sections,
the great
amount of them  released during the core-collapse turns 
 $^{20}$Ne in the outer envelopes of the
collapsing star into  $^{19}$F (Woosley \& Haxton 1988).
\item Novae can
also in principle be $^{19}$F producers (Jos\'e \& Hernanz 1998),
although the yields are still uncertain.  In classical novae the
reaction chain $^{17}$O$(p, \gamma)^{18}$F$(p, \gamma) ^{19}$Ne is the
mechanism involved  in the synthesis of $^{19}$F through  the production
of the short-lived, $\beta^+$-unstable nucleus $^{19}$Ne, which is partially
transferred by convection toward the outer cooler layers of the
envelope, where it decays into $^{19}$F.
\end{enumerate}

Recently, new observational data (related to 49 K giants with
temperatures low enough to show an HF line at 2.3 $\mu$m) by
J{\"o}nsson et al. (2017a) showed that the $\nu$-process is not the
main contributor of fluorine production in the solar vicinity, at
variance with theoretical predictions.
J{\"o}nsson et
al. (2017b) estimated the kinematic probability that stars belong to
the thin disk, thick disk, and/or halo. They showed that the majority
of the observed stars are part of the Galactic thin disk

The observed [F/Fe] vs. [Fe/H] and [F/O] vs. [O/H] ratio both follow
and increasing trend, in contrast to what would expect if the
$\nu$-process was the dominant fluorine nucleosynthesis source in the
solar vicinity, constraining its possible contribution to the cosmic
fluorine budget. This tension between data and the $\nu$-process in the solar neighborhood, is
confirmed  by the observed secondary relation between fluorine and oxygen
in J{\"o}nsson et al. (2017a). This observed secondary behavior can
constrain the stellar models of AGB and W-R stars. Hence,
AGB stars, W-R stars, SNe Ia, novae, and explosive nucleosynthesis in SNe II without
$\nu$ process, are the possible fluorine production sites to be tested
in detail chemical evolution models for the solar neighborhood.
 
In this paper, we present chemical evolution models designed to
reproduce the data by J{\"o}nsson et al. (2017a) in the solar
neighborhood testing different nucleosynthesis prescriptions for the
fluorine yields. We will present results related both to an update
version of the classical ``two-infall'' model introduced by Chiappini
et al. (1997) in which we follow the evolution of the halo-thick disk
and thin disk phases, and to a detailed chemical evolution model which
follows only the thin disk evolution.  The F production sources
considered in this paper are: AGB stars (Karakas 2010), massive stars
(Kobayashi et al. 2006) and W-R stars (Meynet \& Arnould 2000).  
We will also discuss the case  of novae as possible sources for F
production adopting the nucleosynthesis yields by Jos\'e \& Hernanz (1998).

 Previous papers (Meynet \& Arnould 2000, Renda et al. 2004,
   Kobayashi et al. 2011a) have computed fluorine evolution in the
   Galaxy.  For the first time, Meynet \& Arnould (2000) showed that
   W-R stars could be significant contributors to the solar system
   abundance of fluorine using a simple model for the chemical
   evolution. In Renda et al. (2004) the impact of fluorine
   nucleosynthesis, in both W-R and AGB stars, were considered in
   chemical evolution model and concluded that the contribution of W-R
   stars is necessary to reproduce the F abundance in solar vicinity.
   Kobayashi et al. (2011a) found that the main effect of
   $\nu$-process of core-collapse supernovae on the evolution of
   fluorine in the solar neighborhood, is the presence of a plateau at
   high [F/O] values for [O/H] $<$ -1.2 in the [F/O] vs [O/H]
   relation.  As underlined above, this plateau is in contrast with
   the most recent observational data by J{\"o}nsson et al. (2017a).
 Novae have never been included so far in chemical models predicting fluorine.

The paper is organized as follows: in Section 2 we describe the
chemical evolution models for the solar vicinity adopted in this
paper, in Section 3 the nucleosynthesis prescriptions are
described. In Section 4 our results concerning the fluorine abundances
in the solar neighborhood predicted by our chemical evolution models
are reported.  In Section 5, we discuss the effects of the novae
nucleosynthesis on fluorine production.  In Section 6 we draw our main
conclusions.

\begin{table*}[htp]

\caption{The list of the models described in this work in which we considered different contributions for the fluorine production. In this Table we report only the different nucleosynthesis yields adopted for the fluorine.  }
\scriptsize

\label{models}
\begin{center}
\begin{tabular}{c|ccccc}
  \hline
\noalign{\smallskip}

\\
 Models &AGB stars&  SNe Ia& SNe II& Wolf Rayet stars  \\
 &&  &&   \\
   &Karakas (2010)&  Iwamoto et al. (1999)&Kobayashi et al. (2006)
 & Meynet \& Arnould (2000)\\

  \\
\noalign{\smallskip}

\hline
\noalign{\smallskip}

F1& yes &  yes &yes  &no\\
\noalign{\smallskip}
\hline
\noalign{\smallskip}

F2  & no &  yes &yes &no\\
\noalign{\smallskip}
 \hline
\noalign{\smallskip}


F3& yes $\times$ 1.5 &  yes &yes&no\\

\noalign{\smallskip}

 \hline
\noalign{\smallskip}
 F4& yes $\times$ 2 &  yes &yes &no\\
\noalign{\smallskip}

 \hline
\noalign{\smallskip}
 F5& yes $\times$ 3 &  yes &yes &no\\
\noalign{\smallskip}

 \hline
\noalign{\smallskip}
F6 &no &  no &no &yes\\
\noalign{\smallskip}
\hline
\noalign{\smallskip}

F7 & yes &  yes &yes &yes\\

\noalign{\smallskip}

 \hline
\noalign{\smallskip}
 F8& yes &  yes &yes &yes $\times$ 2\\   
\noalign{\smallskip}

 \hline
\noalign{\smallskip}

 F9& yes  $\times$ 2 &  yes &yes   &yes\\

\noalign{\smallskip}

 \hline
\noalign{\smallskip}


\end{tabular}
\end{center}
\label{tab1}
\end{table*}

\section{The chemical evolution models for the solar neighborhood}

In this Section, we describe the main characteristics of the adopted
chemical evolution models for the solar neighborhood in this work:
\begin{enumerate}
\item The  classical ``two-infall'' model of Chiappini et al. (1997, 2001). It is assumed that
the Galaxy formed by two independent infalls of primordial gas. In the
first episode, occurred on short time-scales, the halo-thick disk
components have been formed, while in the second one  the thin disk
was created on longer time-scales.

\item Following  Matteucci \& Fran{\c c}ois (1989)  approach, we use  the ``one-infall'' chemical evolution
model only  for the thin disk evolution in the solar neighborhood.
\end{enumerate}

\subsection{The ``two-infall'' model}
As stated in Section 2, J{\"o}nsson et al. (2017a) data sample  is related to
thick-disk, thin-disk and halo stars. 
For this reason, first  we want to reproduce those data
 adopting the classical ``two-infall'' model capable to trace
the chemical evolution of halo-thick disk and thin disk phases.

 The
Galaxy is assumed to have formed by means of two main infall episodes.
The accretion law for a certain element $i$ at the time $t$  in the solar vicinity is defined as:
\begin{equation}
\mathcal{A}(t,i)=X_{A_{i}}\left[ c_1 \, e^{-t/ \tau_{H}}+ c_2 \, e^{-(t-t_{max})/ \tau_{D}} \right].
\label{a}
\end{equation}
The quantity $X_{A_{i}}$ is the abundance
by mass of the element $i$ in the infalling gas, while
$t_{max}=1$ Gyr is the time for the maximum infall on the thin disk,
$\tau_{H}= 0.8$ Gyr is the time-scale for the creation of the halo
and thick-disk and $\tau_{D}=7$ Gyr is the timescale to build up
the thin disk in the solar neighborhood  (as suggested by
fitting the G-dwarf metallicity distribution).
Here, we assume that the abundances $X_{A_{i}}$ show
primordial gas compositions.
Finally, the coefficients $c_1$ and $c_2$ are obtained by imposing a
 fit to the observed current total surface mass density in the solar neighborhood.
A threshold
 gas density of 7 M$_{\odot}$pc$^{-2}$ in the star formation process
 (Kennicutt 1989, 1998, Martin \& Kennicutt 2001) is also adopted for
 the disk.

The most recent observational data by the Gaia-ESO Survey (Recio-Blanco et al. 2014;
Rojas-Arriagada et al. 2017), APOGEE  (Hayden et al.
2015) and AMBRE  (Mikolaitis et al. 2017) confirmed the existence of
two distinct sequences corresponding to thick and thin disk
stars. In Grisoni et al. (2017) it was shown that the ``two-infall'' model  is  able to perfectly reproduce  the  AMBRE data  in the solar neighborhood, when applied to the thick and thin disks without including the halo.  

\subsection{The ``one-infall'' model}
Because of the fact that the majority of the data presented in J{\"o}nsson et al. (2017a) is supposed to be thin disk stars, we also considered a chemical evolution model which only follows the thin disk.

To reproduce the chemical evolution of the thin disk, we adopt an
updated version of the ``one-infall'' chemical evolution model
presented by Matteucci \& Fran{\c c}ois (1989) using the most recent
nucleosynthesis yield of Romano et al. (2010).
\begin{figure}
\includegraphics[scale=0.7]{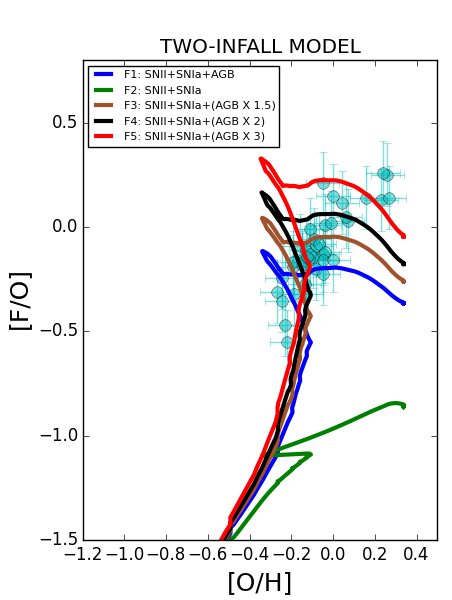}
  \caption{The abundance ratio [F/O] as a function of [O/H] in the
solar neighbourhood for the ``two-infall'' chemical evolution model
adopting different prescriptions for the channels of the fluorine
production. For oxygen we keep Romano et al. (2010, model 15)
prescriptions.  With the blue line we represent the model F1 of
Table \ref{tab1} where fluorine is assumed to be produced by both AGB
stars, SNe Ia and SNe II.  With the green line we have the model F2
(contributions to fluorine production only from SNe Ia and SNe II).
The models F3, F4, F5  in which the AGB yields are multiplied by  factors of 1.5, 2 and 3, respectively (along with SN Ia SN II channels) are drawn with  brown, black, and red lines.
Observational data by   J{\"o}nsson et al. (2017a) are indicated with cyan circles.}
\label{2IM_FO}
\end{figure}
\begin{figure}
\includegraphics[scale=0.7]{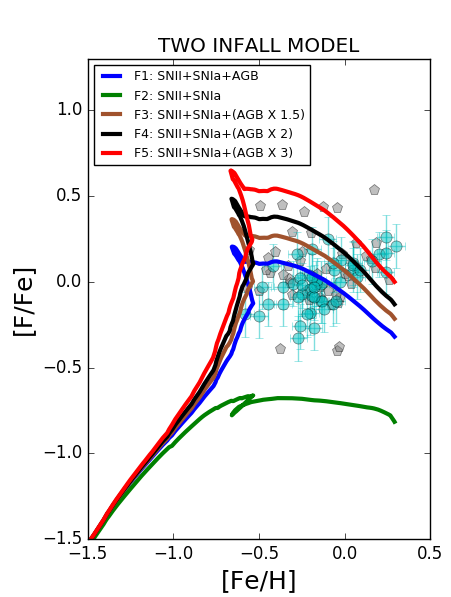}
  \caption{The abundance ratio [F/Fe] as a function of [Fe/H] in the
solar neighbourhood for the ``two-infall'' chemical evolution model
adopting different prescriptions for the channels of fluorine
production. Model lines are the ones described in Fig. \ref{2IM_FO}.
Observational data of   J{\"o}nsson et al. (2017a) are indicated with cyan circles, whereas the data taken by Pilachowski \& Pace (2015) are presented with gray pentagons.}  
\label{2IM_FFE_AGB}
\end{figure} 
The infall rate  in the thin-disk  for  a given
element $i$ at the time $t$, and  in the solar vicinity is
defined as:  
 \begin{equation}\label{infall} 
\mathcal{B}(t,i)= X_{A_i} c_2 \, e^{-\frac{t}{\tau_D}}. 
\end{equation} 
All model parameters are the same as the thin disk phase of the ``two-infall''
model described in Section 2.1.

\section{Nucleosynthesis prescriptions for O, Fe, and F}

In this work our principal  aim is to analyze in detail the contribution of the
different channels for the production of fluorine in the solar
neighborhood,  by comparing the predicted abundance ratios [F/Fe] vs
[Fe/H] and [F/O] vs [O/H] by our chemical evolution models with the data.

 We start our study by considering the
set of nucleosynthesis yields  of the  Romano et al. (2010) best model (their model 15)  for O, Fe and F  (this set of yields has been  adopted  by Brusadin et al 2013, Micali et al. 2013, Spitoni et al. 2016).

In particular, these yields are:

\begin{itemize}
\item For low-and intermediate-mass stars (0.8-8 M$_{\odot}$), we consider
 the metallicity-dependent stellar yields of Karakas
(2010) with thermal pulses. These stars contribute to  fluorine, negligibly to O and give no contribution to Fe.

\item For SNe Ia, the adopted nucleosynthesis prescriptions
are from Iwamoto et al. (1999). These SNe contribute significantly to Fe and negligibly to O and F.

\item  For massive stars (M$>$8 M$_{\odot}$), which are the progenitors of 
either SNe II or hypernovae (HNe), depending on the explosion energy, we assume the 
metallicity-dependent  He, C, N and O stellar yields, 
as computed with the Geneva stellar evolutionary code, which takes
into account the combined effect of mass loss and rotation (Meynet \&
Maeder 2002, Hirschi et al.  2005, Hirschi 2007, Ekstr\"om et
al. 2008). The Kobayashi et al. (2006) yields not including $\nu$-nucleosynthesis are considered for fluorine.
 J{\"o}nsson et al. (2017a) compared their  new observational data with chemical evolution models by  Kobayashi et al. (2011b) in presence of  $\nu$-nucleosynthesis. The model showed  a  plateau at high [F/O] values  at low [O/H] in contrast with the new data. For this reason
we have not taken into account  this kind of nucleosynthesis in this work.

In this paper we test also the effects of W-R yields.
For W-R stars we assume the F production by the models with mass loss by Meynet \& Arnould (2000). These yields do not include oxygen.

Meynet \& Arnould (1993) have suggested that W-R stars could significantly contaminate the Galaxy with $^{19}$F. In their scenario, $^{19}$F is synthesized at the beginning of the He-burning phase from the $^{14}$N
left over by the previous CNO-burning  core,  and  is  ejected  in  the  interstellar  medium when  the  star  enters  its  WC  phase.  Since the mass loss depends on stellar metallicity, the 
$^{19}$F yields  are metallicity-dependent.

For all the elements heavier than oxygen, namely Fe in this study, we assume the
 Kobayashi et al. (2006) yields.

\end{itemize}

\begin{figure}
\includegraphics[scale=0.7]{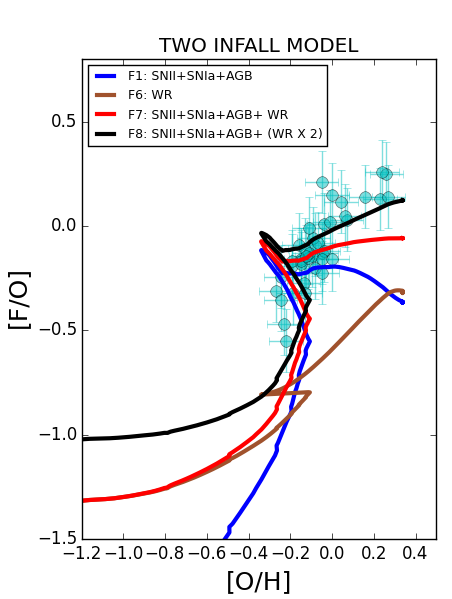}
  \caption{The abundance ratio [F/O] as a function of [O/H] in
the solar neighbourhood for the ``two-infall'' chemical evolution
model testing  different prescriptions for the channels of fluorine
production. 
 With the blue line we represent the model F1 of
Table 1.  With the red line we consider also the Wolf-Rayet yield by Meynet \& Arnould (2000).
Model F6  where only Wolf-Rayet stars  contribute to the fluorine production
 is indicated with the brown  line in this plot. The black
line is  the model F8, similar to the F7 model but  with  the Wolf-Rayet yields multiplied by a factor of 2. Observational data of   J{\"o}nsson et al. (2017a) are indicated with cyan circles.
 }  
\label{WR2_2infall}
\end{figure}

\begin{figure}
\includegraphics[scale=0.7]{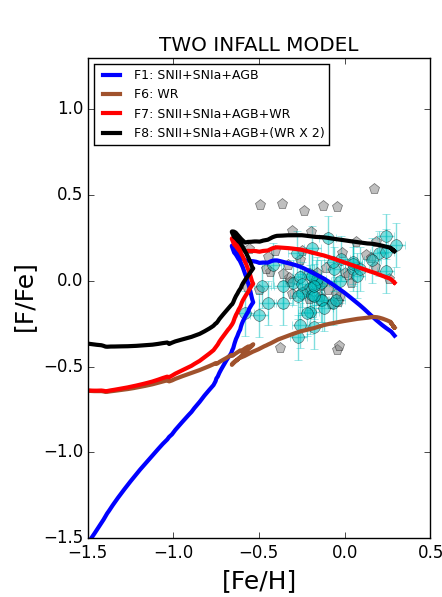}
  \caption{The abundance ratio [F/Fe] as a function of [Fe/H] in the
solar neighbourhood for the ``two-infall'' chemical evolution model
adopting different  fluorine yields. Model lines are the ones described in Fig. \ref{WR2_2infall}.
Observational data of   J{\"o}nsson et al. (2017a) are indicated with cyan circles whereas the data taken by Pilachowski \& Pace (2015) are presented with gray pentagons.}  
\label{WR2_FE_2infall}
\end{figure}

\section{Results: testing different nucleosynthesis prescriptions for fluorine}

In this Section, we show the results related to the different
prescriptions for the fluorine yields with the aim of reproducing the
observational data by J{\"o}nsson et al. (2017a) both for the
``two-infall'' model and for a model where it is studied only the
evolution of the thin disk (``one-infall'' model).  In Table 1 we
present the list of all models tested here varying the fluorine yield
prescriptions.  For each model (F1..F9) it is indicated the presence
of AGB stars with yields by Karakas (2010) in column 2, of Type Ia SNe
with yields by Iwamoto et al. (1999) in column 3, of massive stars
with yields by Kobayashi et al. (2006) in column 4, and W-R stars with
yields by Meynet \& Arnould (2000) in column 5.
As stated above we started our work analyzing the effects of
the nucleosynthesis prescriptions of the best model  by Romano et al. (2010) (their model 15)
on the [F/O] vs. [O/H] and  [F/Fe] vs. [Fe/H] ratios in the light
of the new data by J{\"o}nsson et al. (2017a). We label this model as F1.

The solar values adopted in this work  for  oxygen and iron are the
Asplund (2009) ones, whereas for fluorine the  one computed by Maiorca et al. (2014) is used to be 
coherent with the J{\"o}nsson et al (2017a) work.
\begin{figure}
\includegraphics[scale=0.7]{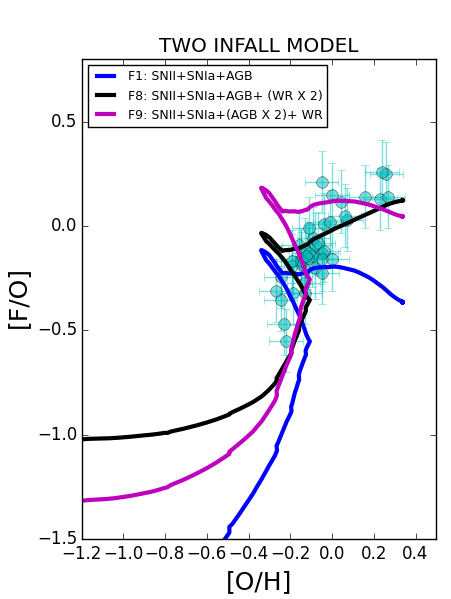}
  \caption{The abundance ratio [F/O] as a function of [O/H] in the
solar neighbourhood for the ``two-infall'' chemical evolution model.
With the blue line we represent the model F1 of
Table 1 where fluorine is assumed to be produced by both AGB
stars, SNe Ia and SNe II. Model, similar to the F7 model but  with  the Wolf-Rayet yields multiplied by a factor of 2 is drawn with the black line. With the magenta  is represented the model F9 where the contribution of SNe II and SNe Ia are considered along with W-R stars. In this model the AGB yields are multiplied by a factor of 2. Observational data of J{\"o}nsson et al. (2017a) are
indicated with cyan circles.} 
\label{F9_2infall}
\end{figure}  
  
We stress that in this work we  analyze in detail only fluorine yields;  for the other elements shown in this
paper (O,Fe), we use the model 15 prescriptions  by Romano et
al. (2010),  which  does not include fluorine though.

\subsection{The ``two-infall'' model results}

 J{\"o}nsson et al. (2017a) compared their new observational data with
 chemical evolution models by Kobayashi et al. (2011a, 2011b) for thick
 and thin disk stars where it was considered the contribution of AGB
 stars and $\nu$ processes to the fluorine production. The main
 problem of those models is that they are not capable  to reproduce the slope of the secondary
 behavior of the new data presented by J{\"o}nsson et al. (2017a).

Here, we show the results related to the ``two-infall'' model of Chiappini
et al. (1997, 2001) updated by Romano et al. (2010) for the chemical
evolution of the solar neighborhood. In Fig. \ref{2IM_FO} we show the
abundance ratio [F/O] as a function of [O/H] in the solar
neighbourhood adopting different prescriptions for the channels of
fluorine production. First, we consider for F and O the
nucleosynthesis yields adopted by the model15 of Romano et al. (2010)
(model F1 in Table 1): the ones of Karakas (2010)  for AGB stars with
thermal pulses, Iwamoto et al. (1999) for Type Ia SNe, and Kobayashi et
al. (2006) for massive stars for F and the Geneva stellar evolutionary
code for O. 

In Fig. \ref{2IM_FO}, model F1 clearly shows the transition between the
halo-thick disk phase and the thin disk. Because of the presence of a
threshold for the surface gas density in the star formation we have a gap in the star formation
history (see Spitoni et al. 2016) in the correspondence of the
beginning of the second gas infall.

At variance with the models of Kobayashi et al. (2011a,b) shown in
Fig. 3 of J{\"o}nsson et al. (2017a), the F1 model shows a decrease of
[F/O] abundance values for [O/H] larger than 0.2 dex.  This behavior
is due to the  different prescriptions for the yield of oxygen we
adopted. In fact, in Romano et al. (2010) the oxygen produced by
massive stars is coming from the Geneva tracks with mass loss and
rotation (Meynet \& Maeder 2002, Hirschi et al.  2005, Hirschi 2007,
Ekstr\"om et al. 2008).
Model F1 is not able  to reproduce the  observational
 trend of J{\"o}nsson et al. (2017a) data   for over-solar values of [O/H], and the main problem is that  we are not capable to reach the high
[F/O] observed values.  

In Fig.  \ref{2IM_FO}, it is also proven that
adopting the Romano et al. (2010) best choice for the yields (their model 15) the main
contribution to the fluorine is given by AGB stars.  
In fact,  model F2 (see Table 1), in which the  fluorine is created only 
by means of
SN Ia and SN II channels, shows a vary small amount  of  fluorine production:   the [F/O] abundance at maximum reaches the value of -0.8 dex.

 However, the AGB yields of F are still uncertain. In fact, the
recent measurements by Indelicato et al. (2017) and He et al. (2017)
of the $^{19}$F$(p, \alpha)^{16}$O reaction, suggest that the computed
yields from AGB-stars should be revised and higher values might be
expected.  Therefore, in Fig. \ref{2IM_FO} the effects of increasing
by hand the contribution by AGB stars are presented.  Multiplying the
yields by factors of 1.5, 2 and 3 (models F3, F4, and F5 in Table 1,
respectively), we have a better agreement with the data.

This
kind of procedure has been widely used in chemical evolution models in
the past.  For example, in Fran{\c c}ois et al. (2004) best model  the Mg
yields related to the massive stars of Woosley \& Weaver
(1995) have been multiplied by a factor of two. More recently, in Maas
et al. (2017) phosphorus yields by Kobayashi et al. (2006)   have been multiplied by a factor of 2.75 with the aim of  
reproducing the [P/Fe] vs [Fe/H] abundance ratios in the solar
neighborhood as  suggested by the new observational stellar data using the chemical evolution model by Cescutti et al. (2012).
From Fig. \ref{2IM_FO} we see that models F4 and F5 are able to:
\begin{itemize}
\item reach the observed high
[F/O] ratio values for  over-solar [O/H] abundances. 
\item trace the observed secondary slope of the data for sub-solar [O/H] values.
\end{itemize}
 The data in Fig.  \ref{2IM_FO}  present a change in the slope of the
  [F/O] vs [O/H] abundance ratios
for values larger than [O/H]=0.
The secondary behavior of the
fluorine is not perfectly reproduced for over-solar values of [O/H], in fact
models F3 and F4  show a decrease of the [F/O] abundance ratio  for [O/H] $>$ 0.

A common problem  of models
F3, F4, and F5, is that the transition between the halo-thick disk
phase and the thin disk has a big impact in the [F/O] vs [O/H] ratios adopting the Romano et al. (2010) model15 set of yields.  We will
show in the next subsection related to the ``one-infall''  results  (the chemical evolution only for the thin disk), that a better agreement with the data is
achieved. We recall here that the majority of the data presented
by J{\"o}nsson et al. (2017a) are parts of the thin disk system.

However, the predicted fluorine solar values are in agreement with 
solar ones of Maiorca et al. (2014) as shown in Table 2.
 In fact, the predicted solar mass fraction
for fluorine  by the model F4 (AGB yields multiplied by a
factor of 2) is  4.91 $\times$ 10$^{-7}$; which compares very well to the value of  4.78 $\times$ 10$^{-7}$  predicted by Maiorca et al. (2014).

\begin{figure}
\includegraphics[scale=0.7]{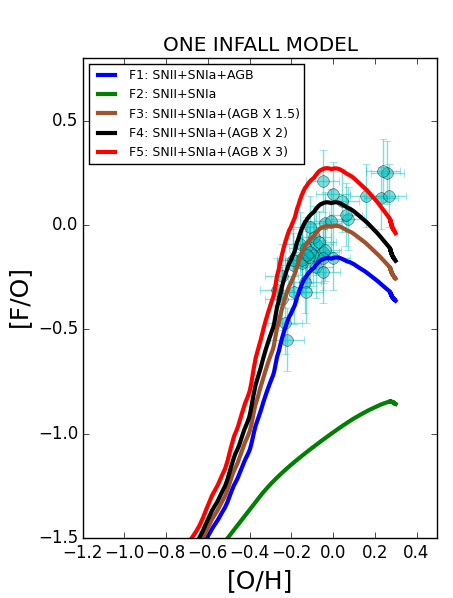}
  \caption{The abundance ratio [F/O] as a function of [O/H] in the
solar neighbourhood for the ``one-infall'' chemical evolution model
adopting different nucleosynthesis prescriptions for the fluorine
production. Model lines are the ones described in Fig. \ref{2IM_FO}.
Observational data of J{\"o}nsson et al. (2017a) are indicated with
cyan circles whereas the data taken by Pilachowski \& Pace (2015)
are presented with gray pentagons. } 
\label{F_O_AGB_VAR_1infall}
\end{figure}

\begin{figure}
\includegraphics[scale=0.7]{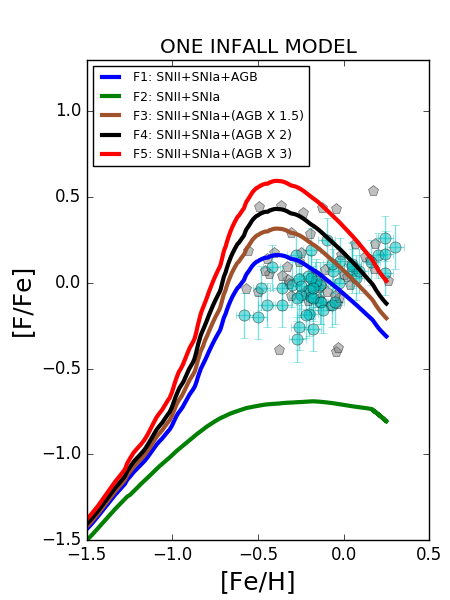}
  \caption{The abundance ratio [F/Fe] as a function of [Fe/H] in the
solar neighbourhood for the ``one-infall'' chemical evolution model.
Model lines are the ones described in Fig. \ref{F_O_AGB_VAR_1infall}. Observational data of   J{\"o}nsson et al. (2017a) are indicated with cyan circles whereas the data taken by Pilachowski \& Pace (2015) are presented with gray pentagons. }  
\label{AGB_var_1infall}
\end{figure}

In Fig. \ref{2IM_FFE_AGB} the same models of the previous Figure are considered but in the terms of  [F/Fe] vs [Fe/H], and we compare our models with the data by J{\"o}nsson et
al. (2017a) along with Pilachowski \& Pace (2015) ones (the
stars in Pilachowski \& Pace 2015 are thin disk, see their Section 2.1).
 If we consider both data
sets the area spread by the data is well covered by the model predictions.
On the other hand, the trend shown by solely  the data of J{\"o}nsson et al. (2017a)  - i.e. the abundance ratio [F/Fe] always increases with [Fe/H] -  is not reproduced by any models presented so far in this paper.

In Fig \ref{WR2_2infall} we present the model results for the
two-infall model where the W-R star yields by Meynet \& Arnould (2000)
for the fluorine are taken into account.  As suggested by Renda et
al. (2004) and J{\"o}nsson et al. (2017a) a possible source for fluorine production are W-R stars. They  might deposit
fluorine into the interstellar medium via their strong stellar winds
(Meynet
\& Arnould 2000; Palacios et al. 2005). Just like in the AGB scenario, the fluorine in W-R winds is produced in
reactions starting from $^{14}$N, including $\alpha$, neutron, and
proton captures. Due to the strong metallicity dependence
of the winds of these stars, a possible fluorine production
through this channel is expected to start first at slightly
sub-solar metallicities and then increase for higher
metallicities (Renda et al. 2004).

\begin{table}[htp]

\caption{The solar values expressed in mass fraction of fluorine predicted the  models reported in Table 1 where different channels for the fluorine production are considered. }  
\scriptsize

\label{models}
\begin{center}
\begin{tabular}{c|ccccc}
  \hline
\noalign{\smallskip}

\\
 Fluorine Solar values   & \multicolumn{2}{c}{Maiorca (2014):  4.78 $\times$ 10$^{-7}$}  \\
 &  &&   \\
 $(X_{F, \mbox{ }\odot})$ &  2 Infall model &1 Infall model  \\
   
  \\
\noalign{\smallskip}

\hline
\noalign{\smallskip}

F1&    3.07 $\times$ 10$^{-7}$ & 3.03 $\times$ 10$^{-7}$ \\
\noalign{\smallskip}
\hline
\noalign{\smallskip}
F2&    9.51 $\times$ 10$^{-8}$ &  9.15 $\times$ 10$^{-8}$\\
\noalign{\smallskip}
\hline
\noalign{\smallskip}

F3&  3.99 $\times$ 10$^{-7}$ &3.95 $\times$ 10$^{-7}$\\
\noalign{\smallskip}
\hline
\noalign{\smallskip}

F4&  4.91 $\times$ 10$^{-7}$ &4.87 $\times$ 10$^{-7}$ \\
\noalign{\smallskip}
\hline
\noalign{\smallskip}

 F5& 6.74 $\times$ 10$^{-7}$ & 6.71 $\times$ 10$^{-7}$ \\
\noalign{\smallskip}
\hline
\noalign{\smallskip}
F6&    3.23 $\times$ 10$^{-7}$ & 3.10 $\times$ 10$^{-7}$\\
\noalign{\smallskip}
\hline
\noalign{\smallskip}
F7&    5.79 $\times$ 10$^{-7}$ & 5.64 $\times$ 10$^{-7}$ \\
\noalign{\smallskip}
\hline
\noalign{\smallskip}
F8&    8.58 $\times$ 10$^{-7}$ &8.30 $\times$ 10$^{-7}$ \\
\noalign{\smallskip}
\hline
\noalign{\smallskip}
F9&    7.62 $\times$ 10$^{-7}$ &  7.48$\times$ 10$^{-7}$ \\
\noalign{\smallskip}
\hline
\noalign{\smallskip}


\end{tabular}
\end{center}
\label{tab1}
\end{table}

\begin{figure}
\includegraphics[scale=0.7]{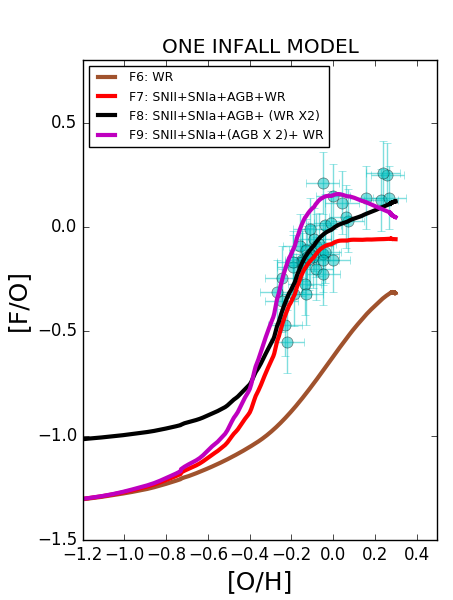}
\caption{The abundance ratio [F/O] as a function of [O/H] in the
solar neighbourhood for the ``one-infall'' chemical evolution model
adopting different prescriptions for the channels of fluorine
production.  The model F6, in which only the contribution for fluorine
production is given by W-R stars, is indicated with the brown line. Model
F7 in which we adopt Romano et al. (2010) yield including W-R stars by
Meynet \& Arnould (2000) is with the red line.  Model F8 (same of
model F7, but with the W-R fluorine yields multiplied by 2, see Table
1) is drawn with the black line. Finally, models F9 (same of model F7,
but with the AGB fluorine yields multiplied by 2, see Table 1) is
labeled with magenta line.  Observational data of J{\"o}nsson et
al. (2017a) are indicated with cyan circles. }
\label{F9_O_H_1infall}
\end{figure} 

 In Fig. \ref{WR2_2infall} we  note that  model F7 which considers the Romano et al. (2010) yield
prescriptions for oxygen,  coupled
with W-R contribution for the fluorine production, is able to increase
the  [F/O] vs [O/H] relation for over-solar values of [O/H] in comparison
with the ``reference'' model F1. This leads to a better agreement with
the data of J{\"o}nsson et al. (2017a), but the slope of the observed
secondary behavior is not well reproduced, and the model shows lower [F/O]
values at larger [O/H].
The predicted solar mass fraction
for fluorine  by the model F7  is  5.79 $\times$ 10$^{-7}$, therefore still in perfect
agreement with Maiorca et al. (2014) value. 

In Fig. \ref{WR2_2infall} we also draw the model F8 (see Table 1 for
model yield detail) where we tested the metal dependent W-R yield by
Meynet \& Arnould (2000) for fluorine multiplied by hand by a factor
of 2 coupled with the Romano et al. (2010) yield.
We note that this model is capable to perfectly reproduce the observed
data. 
  The predicted solar mass fraction for fluorine by the model F8
is 8.58 $\times$ 10$^{-7}$ (see Table 2), definitely a larger value
than the one predicted by Maiorca et al. (2014), but still within a
factor of two.

In summary, from Table 2 we see that the solar
fluorine abundance value reached by the model F7 is in agreement with
the Maiorca et al. (2014) value, whereas the model F8  (W-R fluorine yields
multiplied by a factor of two) shows, as expected, a larger value but
within a factor of two.

We also show the effects of considering fluorine as produced only by
W-R stars (model F6). We note that W-R stars contribute to increase
the fluorine production in the whole range of [O/H]. The main feature
of model F6 is that it shows the secondary behavior slope observed in
the J{\"o}nsson et al. (2017a) stellar sample for [O/H] over-solar
values.  Concerning the [F/Fe] vs. [Fe/H] abundance ratios, in
Fig \ref{WR2_FE_2infall} are reported the same models described above
and shown in Fig.  \ref{WR2_2infall}.

We see that models including the W-R contribution for the fluorine
provide a better fit to the observed data set of J{\"o}nsson et
al. (2017a) and Pilachowski \& Pace (2015) compared to the models
presented in Fig. \ref{2IM_FFE_AGB} without this contribution.

We conclude this Subsection focused on the two-infall model results, presenting
in Fig. \ref{F9_2infall} the effects of varying  AGB fluorine yields by hands
in presence of W-R fluorine contribution. In model F9 (see Table 1) the AGB yields are multiplied by a factor of two   along with W-R yields by Meynet \& Arnould (2000).
In  this way we have a better agreement with data compared to reference 
model F1.

Without any modification of W-R fluorine yields we find a slight
decreasing trend in the [F/O] vs [O/H] relation for over-solar [O/H]
values but still in agreement with  J{\"o}nsson et al. (2017a)
data. Moreover the predicted fluorine solar mass fraction is 7.62 $\times$
10$^{-7}$, therefore in better agreement with the Maiorca et
al. (2014) one.

\begin{figure}
\includegraphics[scale=0.7]{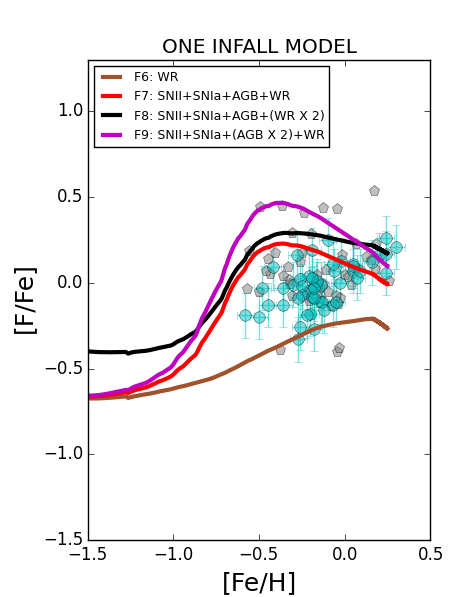}
  \caption{The abundance ratio [F/Fe] as a function of [Fe/H] in the
solar neighbourhood for models F6-F7-F8-F9 applied to the ``one-infall'' chemical evolution model. Model lines are the same of Fig. \ref{F9_O_H_1infall}.
Observational data of   J{\"o}nsson et al. (2017a) are indicated with cyan circles whereas the data taken by Pilachowski \& Pace (2015) are presented with gray pentagons. }  
\label{F9_F_FE_1infall}
\end{figure}

\subsection{The ``one-infall'' model results}
In this Subsection we present the results of chemical evolution models
 for the thin disk of the Galaxy in the solar neighborhood (the one
 infall model).  In Fig. \ref{F_O_AGB_VAR_1infall} models F1, F2, F3,
 F4 and F5 are shown (see Table 1 for model detail).  In
 this Figure it is  evident a ``smoother'' chemical evolution
 compared to the ``two-infall'' case. This is due to the absence of any
 gap in the star formation history which is a peculiar feature of the
 two-infall model (during the transition between the halo-thick disk
 phase) and the thin disk one.

\begin{figure*}
\centering
\includegraphics[scale=0.7]{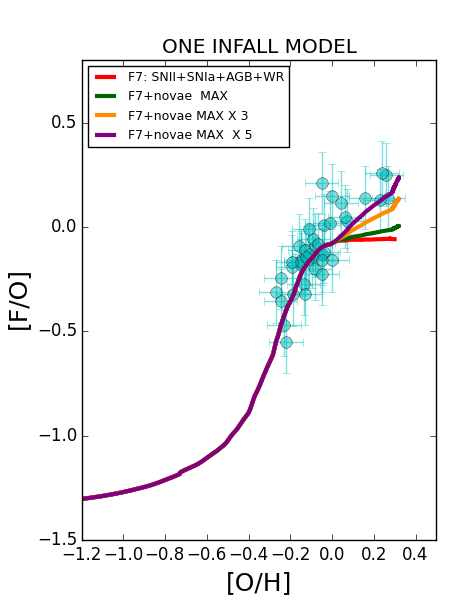}
\includegraphics[scale=0.7]{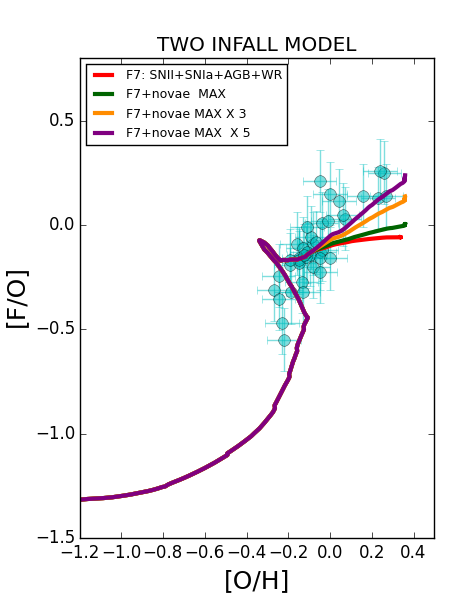}
  \caption{The abundance ratio [F/O] as a function of [O/H] in
the solar neighbourhood for the “one-infall” (left panel) and ``two-infall'' (right panel) chemical evolution
models taking into account the effects of the novae. We consider for fluorine the maximum yield by  Jos\'e \& Hernandez (1998, model ONe7) related to ONe White dwarf with masses of 1.35 M$_{\odot}$). With the  blue line the reference F7 model is reported. The model with the ``maximum'' F novae yield is labeled with the pink line. Models with ``maximum'' F  nova yield multiplied by factor of 5 and 10 are drawn with orange and purple lines, respectively. }  
\label{novae}
\end{figure*}  

As anticipated in Subsection 6.1, the ``one-infall'' model results are
able to better reproduce the J{\"o}nsson et al. (2017a) abundance ratios.
  The same models of
Fig. \ref{F_O_AGB_VAR_1infall} are shown in terms of [F/Fe] vs [Fe/H]
in Fig. \ref{AGB_var_1infall}.

In Fig. \ref{AGB_var_1infall} if we consider both data sets, the area
spread by the data is well covered by model predictions.  On the other
hand, the increase of [F/Fe] for larger values of [Fe/H] shown by the
data of J{\"o}nsson et al. (2017a) alone, is not reproduced by any
model.  Moreover, as it can be seen in Table 2, the solar values
predicted  by the ``one-infall'' model are not so different from the
ones by the two-infall one, although only slightly smaller.

In Fig. \ref{F9_O_H_1infall} we show the effects  on the relations [F/O] vs. [O/H] of including the Wolf-Rayet stars
contribution to the fluorine production on the ``one-infall'' chemical
evolution model for the thin disk  in terms of  [F/O] vs. [O/H]
abundance ratios.  The model which best fits the data is  model F8
which perfectly reproduces the change of slope of the data. In fact, 
 the predicted knee by this model is located  around the  solar value of [O/H],    in agreement with the data.

As for the two-infall  case, the model F8  shows a higher solar value for the  fluorine compared with the Maiorca et al. (2014) value (see Table 2).

Moreover, in Fig. \ref{F9_O_H_1infall} we show the model F9 applied to
the ``one-infall'' chemical evolution, in which we multiplied by a factor
of two the AGB yields for fluorine along with the W-R yield of
Meynet \& Arnould (2000).  Also in this case, the model F9 applied to the
``one-infall'' chemical evolution model leads to a better fit of the data
compared with the case when we considered it for the two-infall
model. Even if this model predicts a decreasing trend for [F/O] vs
[O/H] for over-solar [O/H] values, the model line is within the
observed error bars. As reported in Table 2 this model predicts a
fluorine solar value better in agreement with the Maiorca et
al. (2014) value compared to the model F8.

In Fig. \ref{F9_F_FE_1infall}  models with  W-R fluorine contribution are reported for the  [F/Fe] vs [Fe/H] relation.  Again, if we consider Pilachowski \& Pace (2015) and J{\"o}nsson et al. (2017a) the best model is the F8.  However this model it is not able to  reproduce the trend shown by the  data by J{\"o}nsson et al. (2017a).
In conclusions, as shown  for the two-infall case, we confirm that:
\begin{itemize}
\item Increasing by hand the AGB fluorine yields leads to a better fit of  [F/O] vs [O/H] abundance ratios  for sub-solar values of [O/H], but the secondary behavior at higher [O/H] is not reproduced adopting the nucleosynthesis yields of model15 of Romano et al. (2010).

 \item Models which take into account W-R metallicity dependent yields for fluorine by Meynet \& Arnould (2000)  are  capable to better fit  the observed data.

\item The best models are the F8 and F9 ones.

\end{itemize}

\section{Testing the novae as possible sources for F production}
 In the last section we focus  on the
    effects of nova nucleosynthesis on the
fluorine production. In principle  this source could help to reproduce
the secondary behavior of fluorine  observed by
 J{\"o}nsson et al. (2017a) in the [F/O] vs [O/H] at high [O/H] values.

We tested that the inclusion of the nova yields by Jos\'e \& Hernanz
(1998) on the  model F7 (see Table 1) has a negligible effect on the
chemical evolution of fluorine.  In any case,  as an exercise, we present in
Fig. \ref{novae} the model  with the ``maximum'' fluorine yield by novae, showing the model F7
in the [F/O] vs [O/H] relation, including F produced by novae
originated in ONe white dwarf (W-D) of 1.35 M$_\odot$ (model ONe7 of
Jos\'e \& Hernanz 1998). Although this W-D model leads the maximum
production of F in novae, we are aware that these objects are
extremely rare in nature.  In the same Figure we present model results
in which the fluorine nova  yields are multiplied by factors of 3 and 5,
respectively.

It is quite interesting to note that in this way we are able to trace
perfectly the secondary trend at high [O/H] values.  However, we
remind the reader that nova yields are still very uncertain. It is
worth noting that Li synthesized in a nova outburst as inferred from
recent observations (Tajitsu et al. 2015, Izzo et al. 2015) exceeds by
far the one expected from the theoretical models of Jos\'e \& Hernanz
(1998).  This fact suggests that there is also room for a significant
revision of F yield from novae.

 \section{Conclusions}
In this article we studied in detail the effects of different nucleosynthesis
prescriptions for the fluorine production on the chemical evolution models
for the solar neighborhood with the aim of  reproducing the observational data by J{\"o}nsson et al. (2017a).
Our main conclusions can be summarized as follows:
\begin{itemize}
\item The role played by Wolf-Rayet stars in the fluorine production  seems to be essential to reproduce
the new observed ratios in the solar neighborhood [F/O] vs [O/H] by J{\"o}nsson et al. (2017a). This confirms previous suggestions by Renda et al. (2004).

\item We obtain a better agreement with the observed fluorine abundances when we consider the ``one-infall'' chemical evolution model, relative only to the thin disk.

\item The best ``one-infall'' model  reproducing the observed abundance ratios [F/O] vs [O/H] for the thin disk requires the
nucleosynthesis prescriptions of the model 15  of  Romano et al. (2010) along with the Wolf-Rayet fluorine yields by Meynet \& Arnould (2000)
multiplied  by a factor of two. On the other hand, this model predict a solar value higher than the one of Maiorca et al. (2014).

\item Considering the AGB yields by Karakas et al. (2010)  multiplied by a factor of two along with   Meynet \& Arnould (2000) Wolf-Rayet yields leads to a good fit of the  [F/O] vs [O/H] data and
 predict a  solar value  in agreement with Maiorca et al. (2014) value.

\item Concerning the [F/Fe] vs [Fe/H] relation, the models presented here are in agreement with the collection of data composed by J{\"o}nsson et al. (2017a)  and  Pilachowski \& Pace (2015).
The data by J{\"o}nsson et al. (2017a) show  that [F/Fe] abundances increase with [Fe/H]. This trend is not found by our models for [Fe/H]  values larger than -0.3 dex.   We want to underline that there are still huge uncertainties concerning the nucleosynthesis of F, and this could be the reason of the  discrepancy.

\item More detailed data for fluorine in the solar neighborhood are required
   at low metallicities, i.e for [O/H] values smaller than -0.4 dex, to
   confirm the importance of W-R stars in the fluorine production, as we conclude in this paper. In
   fact, we predict that the inclusion of the fluorine produced  by W-R
   star would affect the [F/O] vs [O/H] ratio also at small [O/H] values
   leading  to  a roughly flat [F/O] ratio for [O/H] smaller that -0.5 dex.

 \item  We also show that the novae, even if their yields are still uncertain could help to better reproduce the secondary behavior of F in the [F/O] vs. [O/H] relation in presence of W-R stars fluorine contribution. 

\end{itemize}

\section*{Acknowledgments}
 We thank the anonymous referee for the suggestions that improved the
paper. E. Spitoni and F. Matteucci thank the financial support by
FRA2016 - University of Trieste. N. Ryde acknowledges support from the
Swedish Research Council, VR (project number 621-2014-5640), Funds the
Royal Physiographic Society of Lund. (Stiftelsen Walter Gyllenbergs
fond and M\"arta och Erik Holmbergs donation), and from the project
grant ``The New Milky'' from the Knut and Alice Wallenberg
foundation. H. J{\"o}nsson acknowledges the support by the Lars Hierta
Memorial Foundation, Helge Ax:son Johnsons stiftelse, and Stiftelsen
Olle Engkvist Byggm\"astare.

\end{document}